\documentclass[twocolumn]{aastex63}

\usepackage{soul}
\usepackage{amsmath}
\usepackage{bm}

\received{April 11, 2022}
\revised{May 27, 2022}
\accepted{June 8, 2022}
\submitjournal{ApJ Letters}

\shortauthors{Goncharov et al.}

\graphicspath{{./}{images/}}

\begin{document}


\title{Consistency of the Parkes Pulsar Timing Array Signal with a Nanohertz Gravitational-wave Background}
\shorttitle{Common versus quasi-common}


\author[0000-0003-3189-5807]{Boris Goncharov}%
 \email{boris.goncharov@me.com}
\affiliation{Gran Sasso Science Institute (GSSI), I-67100 L'Aquila, Italy}
\affiliation{INFN, Laboratori Nazionali del Gran Sasso, I-67100 Assergi, Italy}

\author[0000-0002-4418-3895]{Eric~Thrane}
\affiliation{School of Physics and Astronomy, Monash University, Clayton, VIC 3800, Australia}
\affiliation{ARC Centre of Excellence for Gravitational Wave Discovery}

\author[0000-0002-7285-6348]{Ryan M. Shannon}
\affiliation{Centre for Astrophysics and Supercomputing, Swinburne University of Technology, PO Box 218, Hawthorn, VIC 3122, Australia}
\affiliation{ARC Centre of Excellence for Gravitational Wave Discovery}

\author[0000-0002-7332-9806]{Jan Harms}
\affiliation{Gran Sasso Science Institute (GSSI), I-67100 L'Aquila, Italy}
\affiliation{INFN, Laboratori Nazionali del Gran Sasso, I-67100 Assergi, Italy}

\author[0000-0002-8383-5059]{N.~D.~Ramesh~Bhat}
\affiliation{International Centre for Radio Astronomy Research, Curtin University, Bentley, WA 6102, Australia}





\author{George~Hobbs}
\affiliation{Australia Telescope National Facility, CSIRO, Space and Astronomy, PO Box 76, Epping, NSW 1710, Australia}

\author[0000-0002-0893-4073]{Matthew~Kerr}
\affiliation{Space Science Division, Naval Research Laboratory, Washington, DC 20375-5352, USA}




\author[0000-0001-9445-5732]{Richard~N.~Manchester}
\affiliation{Australia Telescope National Facility, CSIRO, Space and Astronomy, PO Box 76, Epping, NSW 1710, Australia}


\author[0000-0002-2035-4688]{Daniel~J.~Reardon}
\affiliation{Centre for Astrophysics and Supercomputing, Swinburne University of Technology, PO Box 218, Hawthorn, VIC 3122, Australia}
\affiliation{ARC Centre of Excellence for Gravitational Wave Discovery}

\author[0000-0002-1942-7296]{Christopher J. Russell}
\affiliation{CSIRO Scientific Computing, Australian Technology Park, Locked Bag 9013, Alexandria, NSW 1435, Australia}








\author[0000-0001-7049-6468]{Xing-Jiang Zhu}
\affiliation{Advanced Institute of Natural Sciences, Beijing Normal University, Zhuhai 519087, China}

\author[0000-0002-9583-2947]{Andrew~Zic}
\affiliation{School of Mathematical and Physical Sciences, and Research Centre in Astronomy, Astrophysics and Astrophotonics, Macquarie University, NSW 2109, Australia}
\affiliation{Australia Telescope National Facility, CSIRO, Space and Astronomy, PO Box 76, Epping, NSW 1710, Australia}




\begin{abstract}

Pulsar timing array experiments have recently reported strong evidence for a common-spectrum stochastic process with a strain spectral index consistent  with that expected of a nanohertz-frequency gravitational-wave background, but with negligible yet non-zero evidence for spatial correlations required for a definitive detection.
However, it was pointed out by the Parkes Pulsar Timing Array (PPTA) collaboration that the same models used in recent analyses resulted in strong evidence for a common-spectrum process in simulations where none is present.
In this work, we introduce a methodology to distinguish pulsar power spectra with the same amplitude from noise power spectra of similar but distinct amplitudes.
The former is the signature of a spatially uncorrelated pulsar term of a nanohertz gravitational-wave background, whereas the latter could represent ensemble pulsar noise properties.
We test the methodology on simulated data sets.
We find that the reported common process in PPTA pulsars is indeed consistent with the spectral feature of a pulsar term.
We recommend this methodology as one of the validity tests that the real astrophysical and cosmological backgrounds should pass, as well as for inferences about the spatially uncorrelated component of the background.

\end{abstract}


\section{Introduction} \label{sec:intro}

Pulsar timing array (PTA) experiments pursue the goal of detecting nanohertz-frequency gravitational waves through temporal and spatial cross-correlation of pulse arrival times from millisecond radio pulsars.
The primary target sources of such signals are coalescing supermassive binary black holes separated by less than $\sim 0.1$ pc.
Nanohertz gravitational waves produce correlations in the \textit{timing residuals} between the measured arrival times and the arrival times predicted by the deterministic pulsar timing models~\citep{EdwardsHobbs}.
Recent searches for the stochastic gravitational wave background by NANOGrav\footnote{The North American Nanohertz Observatory for Gravitational Waves~\citep{McLaughlin2013}}~\citep{ArzoumanianBaker2020}, PPTA\footnote{The Parkes Pulsar Timing Array~\citep{ManchesterHobbs2013}}~\citep{GoncharovShannon2021}, EPTA\footnote{The European Pulsar Timing Array~\citep{DesvignesCaballero2016}}~\citep{ChenCaballero2021} and IPTA\footnote{The International Pulsar Timing Array~\citep{VerbiestLentati2016}, a consortium of all pulsar timing arrays across the world}~\citep{AntoniadisArzoumanian2022} reported evidence for the ``common-spectrum process'', the same power-law component in Fourier spectra of timing residuals.
The spatial correlations necessary to claim a detection originate from the so-called \textit{Earth term} of the gravitational wave background~\citep{HellingsDowns1983}.
The \textit{pulsar term} of the background arises from the passage of gravitational waves near the pulsars and only manifests as a spatially uncorrelated process with the same spectrum of temporal correlations in all of the pulsars.
It is expected that evidence for the common-spectrum process, to which both of the terms contribute, would precede the detection of the gravitational wave background~\citep{RomanoHazboun2021,PolTaylor2021}.
\cite{GoncharovShannon2021}, on the other hand, pointed out that the methodology employed by~\cite{ArzoumanianBaker2020} does not allow to distinguish between common and similar noise processes in pulsars.
So, it is unclear if the recent searches have detected a {\em bona-fide} gravitational wave background.

One might ask why do we need to precisely determine the degree of similarity in pulsar spectra if the detection mostly depends on spatial correlations anyway?
There are several reasons.
Noise processes with similar and not common noise spectra could arise from e.g. spin noise, stochastic irregularities in rotation of the pulsars~\citep{ShannonCordes2010}.
In fact, the spin noise model discussed in~\cite{MeyersO'Neill2021} suggests a spectral index of timing residuals to be $\gamma =4$  (where the power spectral density is modelled as $P(f) \propto f^{-\gamma}$), which would be difficult to distinguish from the nanohertz gravitational wave background expected from binary supermassive black holes, $\gamma = 13/3$, with current PTAs~\cite[see e.g. Figure 13 in][to inspect measurement uncertainties]{RenziniGoncharov2022}.
Empirical models for spin noise in millisecond pulsars predict timing noise having spectral indices and amplitudes similar to that expected of the gravitational wave background. 
If the reported signal in PTAs is not common and thus not of a gravitational-wave origin, it could have interesting implications for spin noise and hence for neutron star physics~\citep{MelatosLink2014}.
Furthermore, the amplitude of the common-spectrum process is in tension with several predictions for the stochastic background amplitude.
Recent work by~\cite{Izquierdo-VillalbaSesana2022} suggests that the gravitational wave background of the same strain amplitude as the common-spectrum process is challenging to produce in theory given the constraints from the quasar bolometric luminosity functions or the local black hole mass function.
\cite{Casey-ClydeMingarelli2022} find the local number density of supermassive binary black holes inferred from the amplitude of the common-spectrum process, assuming it to originate from the stochastic background, to be five times larger than theoretical predictions.
So, even if PTAs are observing hints of a gravitational wave background, inferences based on the common-spectrum process might be contaminated by pulsar-intrinsic noise, and it is important to clarify to what degree it is true.
Such inferences can be promising because the constraints on the background amplitude from inter-pulsar correlations lag behind the ones based on spatial auto-correlations for most pulsar timing arrays~\citep{PolTaylor2021}.

Whereas previous analyses of time-correlated noise in pulsars were based on identifying~\citep[e.g.][]{LentatiShannon2016,GoncharovReardon2021} or modelling~\citep{CaballeroLee2016,GoncharovZhu2020,ChalumeauBabak2022} noise power spectra, here we focus on modelling Bayesian priors using the second PPTA data release \citep[PPTA-DR2,][]{KerrReardon2020}.
The methodology is described in Section~\ref{sec:method}. 
In Section~\ref{sec:results} we outline the results and we summarise the conclusions in Section~\ref{sec:conclusions}.

\section{Methodology} \label{sec:method}

\subsection{Inference of the common-spectrum process} \label{sec:method:standard}

The timing residuals comprise a number of stochastic processes.
Those that are not due to gravitational waves are considered as noise.
Temporally correlated processes are called \textit{red}, whereas processes without temporal correlations are referred to as \textit{white}.
The power spectral density of red processes is usually assumed to be a power law:
\begin{equation}\label{eq:powerlaw}
    P(f|A,\gamma) = \frac{A^2}{12 \pi^2} \bigg(\frac{f}{\text{yr}^{-1}}\bigg)^{-\gamma} \text{yr}^3,
\end{equation}
where the amplitude $A$ is in the units of strain amplitude of the stochastic gravitational wave background at $f = \text{yr}^{-1}$ and $-\gamma$ is a spectral index.
The frequency in Fourier spectra of fluctuations in timing residuals and concurrently a frequency of a gravitational wave that would have induced these fluctuations is denoted $f$.
To clarify, for white noise $P$ is a constant, it does not depend on $f$.
Red process spectra are modelled at $n_c$ harmonically related frequencies that are multiples of the reciprocal of the observation span.
Values of $n_c$ as well as priors for $A$ and $\gamma$ and the list of noise terms found in PPTA DR2 were published in~\cite{GoncharovShannon2021}.
Additional details on the noise models in PPTA DR2 are outlined by~\cite{GoncharovReardon2021}.
All noise, timing model parameters and signals of interest are modelled using the multivariate Gaussian likelihood \cite[][]{LentatiShannon2014,ArzoumanianBrazier2016,TaylorLentati2017} and Bayesian posterior sampling.
Without accounting for spatial correlations, the total PTA likelihood is a product of individual pulsar likelihoods.
In particular, for a common-spectrum process with $A_\text{c}$ and $\gamma_\text{c}$, the total PTA likelihood is:
\begin{equation}\label{eq:signalstandard}
    {\cal L}(\bm{d} | \bm{\theta}, A_\text{c}, \gamma_\text{c}) = \prod_{k=1}^N {\cal L}(\bm{d}_k | \bm{\theta}_k, A_\text{c}, \gamma_\text{c}),
\end{equation}
where $\bm{\theta} = (\bm{\theta}_1, ..., \bm{\theta}_N)$ are parameters of models that describe data of individual pulsars $\bm{d} = (\bm{d}_1,...,\bm{d}_N)$, including pulsar-intrinsic ``spin'' noise parameters $A_k,\gamma_k$ and parameters of other noise terms which are not of interest to us for the purpose of this work.
Some of these parameters are marginalized over analytically~\cite[][]{2009MNRAS.395.1005V} and others numerically\footnote{These parameters are fit for but not presented in our results. Posterior samples for $\bm{\theta}_k$ are reweighted in target and proposal likelihoods, and the total likelihood is generally independent of them unless otherwise specified.}, so $\bm{\theta}$ is omitted from the following equations.
Both for the common-spectrum process found in PPTA DR2 and for a classical model of the stochastic gravitational wave background from circular supermassive black hole binaries, $\gamma_\text{c} = 13/3$. 
We will fix $\gamma_\text{c}$  at this value throughout our analysis.
We provide additional remarks on the data and analysis in Appendix~\ref{sec:notes}.

\subsection{Importance sampling for pulsar timing arrays} \label{sec:method:importance}

When it is computationally challenging to evaluate a likelihood or to include several measurements in one likelihood, many data analyses resort to the so-called \textit{importance sampling}~\citep[Chapter 10 in][]{gelman1995bayesian, PayneTalbot2019}.
The idea is that the analysis is first carried out assuming a \textit{proposal distribution} -- a likelihood or a prior which is easy to evaluate.
Next, posterior samples obtained from this first step are used to evaluate the \textit{target distribution} -- a likelihood or a prior that represents the model we are ultimately interested in.
Finally, if proposal samples are collected from subsets of a total data set, they can be combined into a single likelihood through the procedure known as posterior recycling~\citep{ThraneTalbot2019}.
To sum up, importance sampling revolves around reweighing of likelihoods and priors.

Let us represent our signal likelihood given by Equation~\ref{eq:signalstandard} through target likelihoods with the common-spectrum process, ${\cal L}(d_k | A_{k,j}, \gamma_{k,j}, A_\text{c}, \gamma_\text{c})$, and the proposal likelihood without the common-spectrum process, ${\cal L}(d_k | A_{k,j}, \gamma_{k,j})$.
Both the proposal and the target likelihoods will represent individual pulsars, whereas the total signal likelihood includes contributions from all pulsars:
\begin{equation}\label{eq:signal}
    {\cal L}(\bm{d} | A_\text{c}) =
    \prod_k^N {\cal Z}(d_k) \frac{1}{n_k} \sum_{j=1}^{n_k}
    \frac{
    {\cal L}(d_k | A_{k,j}, \gamma_{k,j}, A_\text{c})
    }{
    {\cal L}(d_k | A_{k,j}, \gamma_{k,j})
    } .
\end{equation}
Here, the sum is over the $n_k$ fiducial posterior samples $j$ for pulsar $k$, generated for the proposal distribution.
The product is over pulsars.
The posterior samples include $(A_k, \gamma_k)$, the amplitude and the spectral index of the red noise for pulsar $k$.
We use these fiducial samples as the ``proposal distribution'' in order to explore a more complicated likelihood (the ``target distribution'').
The Bayesian evidence $\mathcal{Z}(d_k)$ is the integral of the (proposal) likelihood over the prior.
Importance sampling is similar to the factorized likelihood approach~\citep{TaylorSimon2022}, where the amplitude of the common-spectrum process is obtained through the multiplication of posterior distributions obtained in analyses of individual pulsar data.

\subsection{Common versus quasi-common} \label{sec:method:cpqcp}

In the case of a common-spectrum process, which could originate from the gravitational wave background, nature provides us with one $A_\text{c}$ in all pulsars.
In the language of a hierarchical Bayesian statistics~\citep[Chapter 5 in][]{gelman1995bayesian}, $A_{\text{c},k}$ for a pulsar $k$ is drawn from a delta function distribution, described by a \textit{hyper-parameter} that determines the position of a delta function distribution along possible values of $A_\text{c}$.
The standard null hypothesis is that there is no common-spectrum process and pulsar data sets are described by individual pulsar noise.
In this work, we propose an alternative null hypothesis where $A_{\text{c},k}$ are drawn from a Gaussian distribution described by hyper-parameters $\sigma_A,\mu_A$ and not from the delta function distribution described by $A_\text{c}$.
So, in Equation~\ref{eq:signal} we introduce multiple possible values of $A_{\text{c},k}$ and marginalize over the likelihood of such a noise process over the Gaussian\footnote{We assume a truncated Gaussian distribution because, in reality, our measurements are truncated by uniform prior boundaries that are used to obtain proposal posterior samples.} prior $\pi(A_{\text{qc},k} | \mu_A, \sigma_A)$ with hyper-parameters $\mu_A$ and $\sigma_A$ being the mean and the standard deviation of the Gaussian prior, respectively:
\begin{widetext}
\begin{equation}\label{eq:noise}
    {\cal L}(\bm{d} | \mu_A, \sigma_A) = \prod_k^N {\cal Z}(d_k) \frac{1}{n_k} 
    \sum_{j=1}^{n_k}
    \frac{\int {\cal L}(d_k | A_{k,j}, \gamma_{k,j}, A_{\text{qc},k})
    \, \pi(A_{\text{qc},k} | \mu_A, \sigma_A) dA_{\text{qc},k}}{{\cal L}(d_k | A_{k,j}, \gamma_{k,j}, A_{\text{qc},k})}.
\end{equation}
\end{widetext}
We provide a derivation in the Appendix~\ref{sec:maths}.
For $\sigma_A =0$, $\pi(A_{\text{qc},k} | \mu_A, \sigma_A)$ reduces to the delta function.
Therefore, the likelihood ${\cal L}(\bm{d} | A_\text{c})$ in Equation~\ref{eq:signal} is a subset of the likelihood ${\cal L}(\bm{d} | \mu_A, \sigma_A)$ in Equation~\ref{eq:noise}.
We refer to noise processes with $\sigma_A \neq 0$ as \textit{quasi-common}, which means that noise spectra in timing array pulsars are similar but not common.
The measurement of $\sigma_A$ and $\mu_A$ described in this Section is also applicable to modelling pulsar spin noise alone, which is broadly-distributed in $A_k$.
Instead, here we only infer parameters on the second common term $A_{\text{qc}}$ to map the measurements directly to the result of the gravitational wave search with PPTA DR2~\citep{GoncharovShannon2021}.
The principle of the quasi-common noise model is similar to the dropout analysis~\citep[see Figure 9 in][]{ArzoumanianBaker2020}, it quantifies how constraints on the common-spectrum process with one of the pulsars are consistent with constraints from the rest of the pulsars.
As Bayes factors, high dropout factors for pulsars support the common-spectrum process, whereas low dropout factors that approach zero illuminate pulsars that do not have the same spectra as the others.

We calculate the integral over $A_{\text{qc},k}$ for every evaluation of the likelihood ${\cal L}(\bm{d} | \mu_A, \sigma_A)$ in Equation~\ref{eq:noise} the following way.
First, we pre-compute ${\cal L}(d_k | A_{k,j}, \gamma_{k,j}, A_{\text{qc},k})$ on a grid of $A_{\text{qc},k}$ for $n_k$ posterior samples for $N$ pulsars.
Next, for every parameter sample $(\mu_A, \sigma_A)$, we evaluate the prior for the grid of $A_{\text{qc},k}$, multiply it by the pre-computed likelihood ${\cal L}(d_k | A_{k,j}, \gamma_{k,j}, A_{\text{qc},k})$, and evaluate the integral over the product numerically.
Because parameter estimation is traditionally performed on the (base-10) logarithm of the red process amplitude, in practice we measure $\mu_{\log_{10}A}$ and $\sigma_{\log_{10}A}$ instead of $\mu_A$ and $\sigma_A$.

\begin{figure*}
    \includegraphics[width=0.49\textwidth]{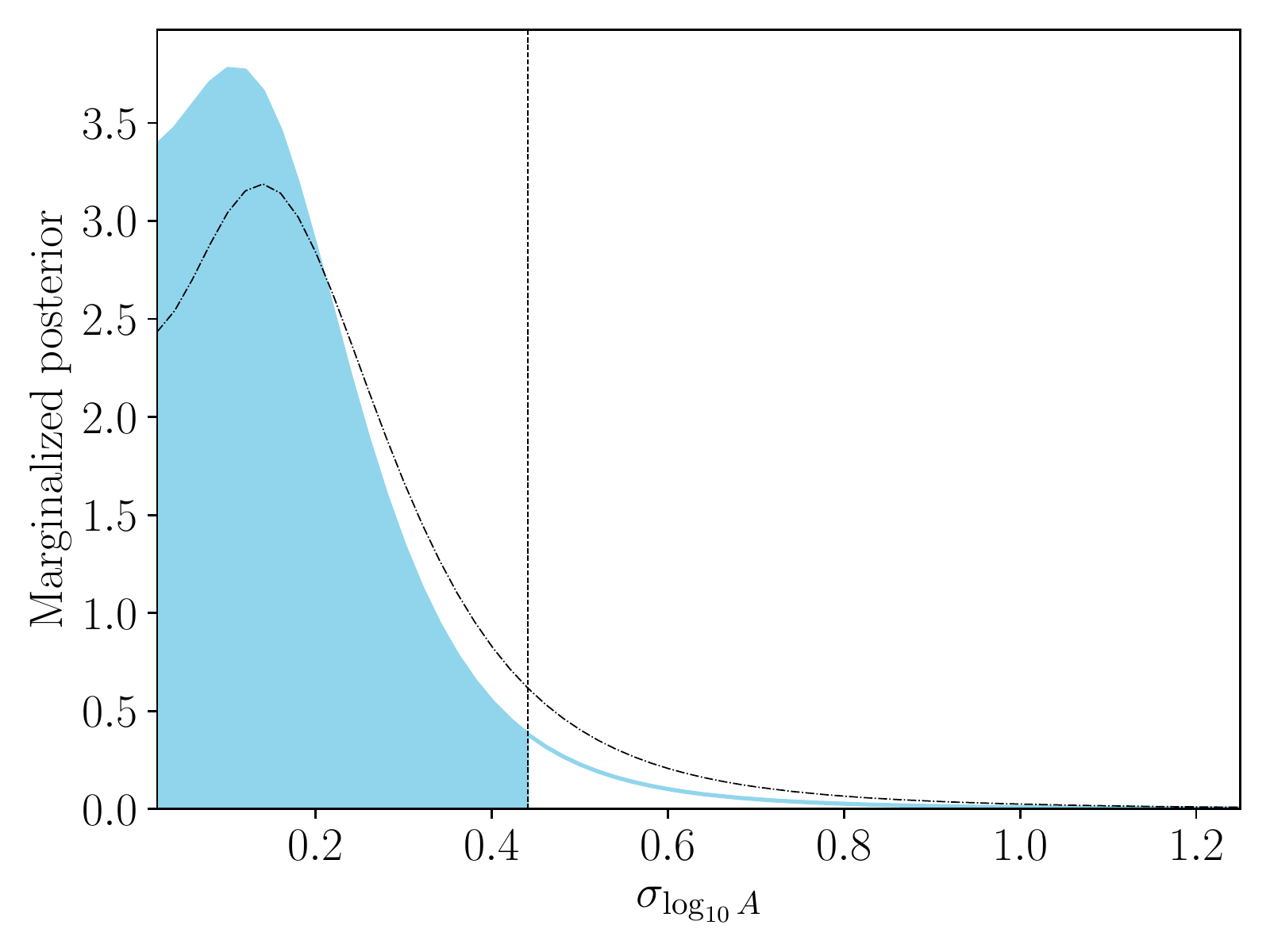}
    \includegraphics[width=0.49\textwidth]{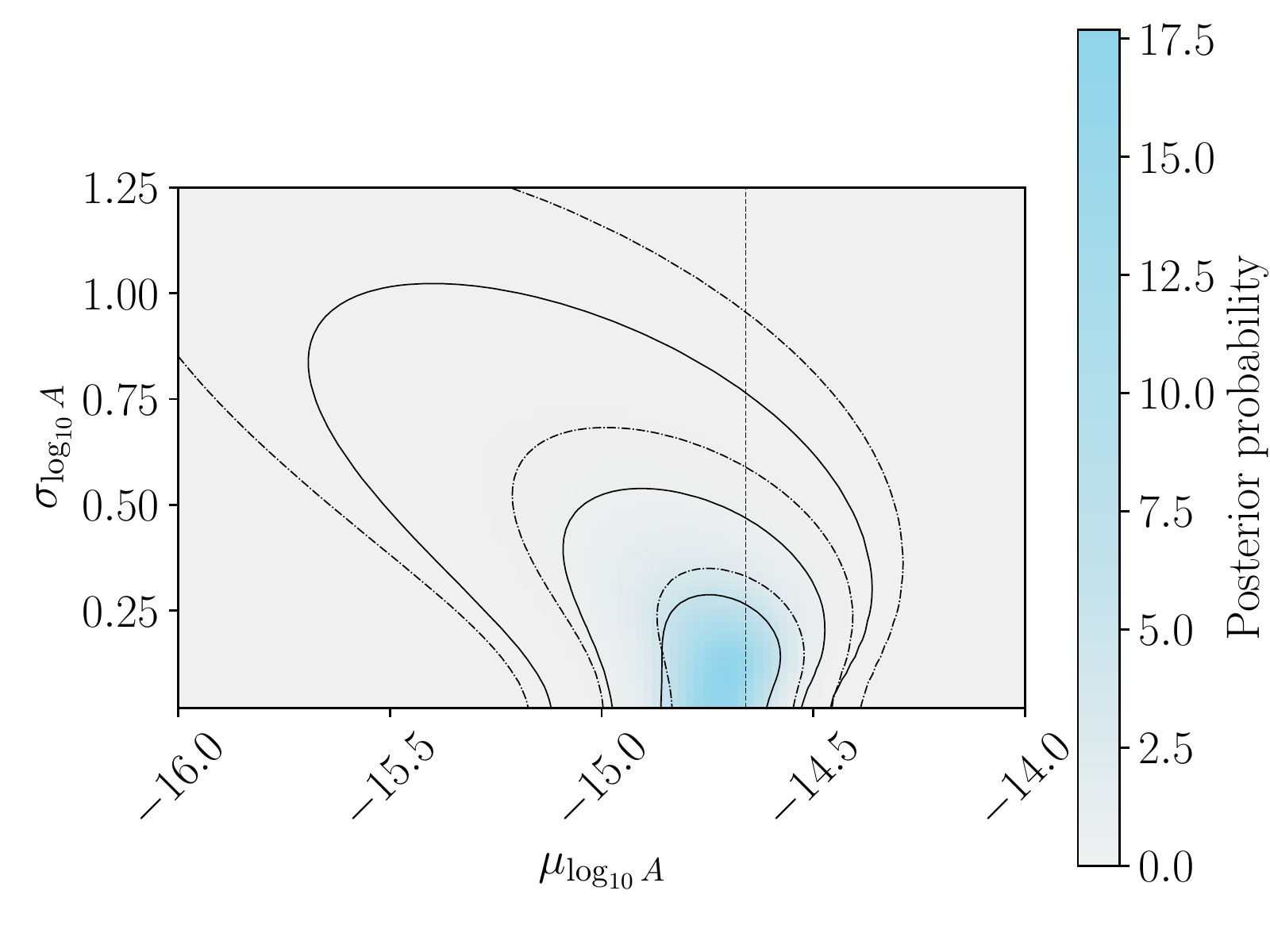}
    \caption{Comparison of common-spectrum and quasi-common spectrum processes in the PPTA DR2.
    Left: marginalized posterior for $\sigma_{\log_{10}A}$. The value is consistent with zero, which means that the common-spectrum hypothesis holds for the PPTA data.
    Vertical dashed line is the upper limit on $\sigma_{\log_{10}A}$ at 95-\% credibility.
    Right: full posterior for $\mu_{\log_{10}A}$ and $\sigma_{\log_{10}A}$. Vertical dashed line corresponds to the measurement of $\log_{10}A$ of the common-spectrum process in PPTA DR2~\citep{GoncharovShannon2021}. Our inference of the common-spectrum process amplitude is consistent with the previous measurement considering generalization of the model and the use of likelihood reweighting.
    Three closed lines correspond to the standard $1$-, $2$- and $3$-$\sigma$ levels.
    In both panels, the dash-dotted lines show the measurements performed without three PSRs with special noise properties discussed in Section~\ref{sec:results}: J0437$-$4715, J0711$-$6830, and J1643$-$1224. Removing these pulsars from the analysis increases uncertainty levels, as expected.
    }
    \label{fig:sigma}
\end{figure*}

\section{Results} 
\label{sec:results} 

We measure $\mu_{\log_{10}A}$ and $\sigma_{\log_{10}A}$ in PPTA DR2 \cite[][]{KerrReardon2020}, with a particular emphasis on $\sigma_{\log_{10}A}$ which distinguishes a common-spectrum process from a quasi-common noise process, as pointed out in Section~\ref{sec:method:cpqcp}.
We find Savage-\cite{dickey1971}
natural log Bayes factor in favor of $\sigma_{\log_{10}A} = 0$ against other $\sigma_{\log_{10}A}$ within the uniform prior to be $-0.16$.
This means that $\sigma_{\log_{10}A}$ is consistent with zero.
We further calculate an upper limit on $\sigma_{\log_{10}A} < 0.44$ at 95\% credibility.
Thus, we demonstrate that the data are more consistent with a common-spectrum process hypothesis than the extended quasi-common process model. 
We measure $\mu_{\log_{10}A} = -14.71^{+0.08}_{-0.14}$, which is consistent with $\log_{10}A=-14.66 \pm 0.07$ found in~\cite{GoncharovShannon2021}.
The result of parameter estimation is provided in Figure~\ref{fig:sigma}.
We note that the maximum \textit{a posteriori} value of $\sigma_{\log_{10}A}$ differs from zero, which may represent a noise fluctuation.

We checked that the result is robust to the exclusion of three pulsars with particular noise properties: PSRs J0437$-$4715, J0711$-$6830 and J1643$-$1224.
PSRs J1643$-$1224 and J0711$-$6830 show spin noise with unusual spectral indices and so we consider them population outliers.
These two pulsars also do not contribute significantly to the common-spectrum process based on the dropout analysis~\cite[Fig. 2 in][]{GoncharovShannon2021}, which is why it is acceptable to exclude them.
PSR J0437$-$4715 shows excess noise that could, in principle, marginally affect measurement of the common-spectrum process, and that does affect the inference of spatial correlations~\citep[see the left panel of Figure 3 in][]{GoncharovShannon2021}.

Additionally, we tested our methodology using simulated data sets. The  details of the simulations can be found in the Appendix~\ref{sec:simqcp2}.
One important issue that we identified is that it is essential to include the common-spectrum process in the proposal likelihood, not only in the target likelihood when testing  the quasi-common noise hypothesis. 
If it is excluded, the measurement of $\sigma_{\log_{10}A}$ will always peak at zero, even if data sets were simulated with e.g. $\sigma_{\log_{10}A} = 2$.
Th{is} is because the proposal likelihood does not sufficiently match the target likelihood.
Non-inclusion of the separate common-spectrum process term will prevent accumulating posterior support in two areas of the parameter space that correspond to the spin noise and the common-spectrum process, respectively.
Nevertheless, our methodology is very sensitive to sub-threshold red-noise contributions, which is generally true for all Bayesian inference that combines data from multiple measurements \cite[e.g.,][]{GoncharovZhu2020}.
In particular, even though two power law processes -- (quasi-)common-spectrum process and pulsar-intrinsic spin noise -- are not individually resolved in single pulsars, Bayesian inference with the total data set distinguishes them when models match the simulation.

\section{Conclusions}
\label{sec:conclusions} 

We have shown that a pulsar timing array data set, in our case PPTA DR2, contains a red process that has spectra consistent with stochastic time-series realizations with the single power-law amplitude by measuring $\sigma_{A}$ consistent with zero.
Note, this conclusion is (a) made under the assumption that the red process power-law index is $13/3$, which is supported by the PPTA data, but (b) not yet applicable to all available PTA data sets. This discrepancy can potentially be resolved through further data combinations, for example by the IPTA.
The identified common-spectrum process in the PPTA DR2 could therefore be the spatially-uncorrelated component of the stochastic gravitational wave background.
We expect the variance in measured pulsar spectra from spatial correlations of the pulsar term to exceed that from different time-series realizations of the Earth term, and thus the uncertainty in $\sigma_A$ to be dominated by the pulsar term. A more detailed investigation of the contribution of both terms to $\sigma_A$ is a subject of a follow-up study.

The methodology of previous nanohertz gravitational wave searches might have led to identifying a common-spectrum process when none is present in the data~\citep{GoncharovShannon2021}.
Specifically, incorrect conclusions can be caused by a mismatch between uniform noise priors on $A$ and $\gamma$ in the models and the clustering of these parameters in real data.
In particular, for the case of a single power-law red noise term in pulsars, evidence for the common-spectrum process disappears when the distribution of true parameters of such noise processes matches the Bayesian priors.
The demonstration of the above two points is to be provided by A.~Zic et al. (in prep.) simulations where the apparent common-spectrum process arises in a variety of scenarios that only contain pulsar-intrinsic spin noise, whereas the spurious emergence of spatial correlations in such data sets is very unlikely.
Determining that the identified noise process is quasi-common could either mean that it is not the gravitational wave background or that the assumed uniform prior distributions for pulsar-intrinsic noise need to be replaced by more realistic priors.
The latter may inform about stochastic irregularities in rotational properties of neutron stars and thus about neutron star physics~\citep{MelatosLink2014}.

Neither our test nor any other test will be able to confirm the common-spectrum hypothesis because it is a matter of measuring the width of the distribution, and we would need an infinitely small measurement uncertainty to rule out all $\sigma_A$ other than zero.
Thus, the techniques we introduce allow to perform a consistency test and to infer common noise properties, but cannot be used to detect the stochastic background.
Once the gravitational wave background is detectable through spatial correlations, further modelling of the priors of pulsar-intrinsic spin noise with the techniques we outlined will be required to disentangle these terms.
Further simulation study will be useful to test the fidelity of the presented methods under different conditions.

We would also like to point out that it is challenging to extend the methodology to allow $\gamma$ to vary and thus to measure $\mu_\gamma$ and $\sigma_\gamma$ simultaneously with $\mu_A$ and $\sigma_A$ because the accuracy of numerical integration on a grid will decline for a higher-dimensional problem and the computational burden will significantly increase.
We foresee several other modifications of the outlined methodology.
For example, one could model the distribution of common-process amplitudes $A$  linearly (rather than logarithmically), as well as to fix the amplitude and fit for the distribution of power-law indices.
Moreover, one could apply our approach to fitting the distribution of pulsar-intrinsic noise amplitudes and power-law indices, which would result in larger $\sigma_A$ and $\sigma_\gamma$.
Furthermore, in case of finding stronger evidence for quasi-common noise with $\sigma_A > 0$, it is possible to test a range of functional forms outlining the distribution of $A$ other than the Gaussian distribution we have assumed.

From the perspective of future validation of the detection of the stochastic background, the tests we propose are complementary to other work.
The dropout analysis in~\cite{ArzoumanianBaker2020} is developed to find outliers in the common-spectrum process.
\cite{TaylorGair2013} introduced interpolant-based modeling for spatial correlations, which is important to ensure that pulsars exhibit precisely the Hellings-Downs correlations and not something else.
\cite{JohnsonVigeland2022} examine a bias from using a finite number of pulsars.
\cite{TaylorSimon2022} develop a factorized likelihood approach for cross-validation of measurements between subsets of timing array pulsars.
\cite{Romero-ShawThrane2022} review several other tests to spot incorrectly specified models.
Because, as outlined by~\cite{GoncharovReardon2021}, the background noise in pulsar timing arrays is often not white and Gaussian as assumed by current models and simulations, we suggest to study the effect of this noise on gravitational wave searches.


\section*{Acknowledgements} \label{sec:acknowledgements}
We thank Matthew Miles, Paul Baker, Stephen Taylor and Sarah Vigeland for useful comments.
This work has been carried out using the Parkes Pulsar Timing Array, which is part of the International Pulsar Timing Array.
Murriyang, the Parkes radio telescope is part of the \href{https://ror.org/05qajvd42}{Australia Telescope}, which is funded by the Commonwealth Government for operation as a National Facility managed by CSIRO. This paper includes archived data obtained through the CSIRO Data Access Portal (\href{http://data.csiro.au}{data.csiro.au}).
BG is supported by the Italian Ministry of Education, University and Research within the PRIN 2017 Research Program Framework, n. 2017SYRTCN.
RMS acknowledges support through Australian Research Council Future Fellowship FT190100155.
Part of this work was undertaken as part of the Australian Research Council Centre of Excellence for Gravitational Wave Discovery (CE170100004).
Work at NRL is supported by NASA.
\software{\textsc{pypolychord}~\citep{HandleyHobson2015}, \textsc{ptmcmcsampler}~\citep{EllisvanHaasteren2019}, \textsc{enterprise}~\citep{EllisVallisneri2019}, \textsc{bilby}~\citep{AshtonHubner2019}, \href{https://github.com/bvgoncharov/enterprise_warp}{github.com/bvgoncharov/enterprise\_warp}, \href{https://github.com/bvgoncharov/ppta_dr2_noise_analysis}{github.com/bvgoncharov/ppta\_dr2\_noise\_analysis}.}

\appendix

\section{Notes on data and analysis}
\label{sec:notes}

We emphasise that the sources of noise, especially, the red-noise terms, should be modelled, to ensure the correctness of the analysis.
The second data release of the PPTA contains several sources of red noise which were found and identified in~\cite{GoncharovReardon2021} based their attribution to telescope observing band or system as well as on their radio frequency dependence.
\cite{GoncharovReardon2021} also identified non-stationary sources of noise which affect red noise measurements if not modelled.
Incorrect modelling of pulsar timing model parameters such as spin frequency derivatives or instrumental phase jumps can also appear as red noise.
Pulsar timing models for PPTA DR2 in coordination with the noise analysis were performed by~\cite{ReardonShannon2021}.
The data set and the code is available at~\href{https://github.com/bvgoncharov/ppta_dr2_noise_analysis}{github.com/bvgoncharov/ppta\_dr2\_noise\_analysis}.

Parameter estimation and Bayesian evidence evaluation for individual pulsars is performed with nested sampling~\citep{skilling2006} implemented in~\textsc{pypolychord} by~\cite{HandleyHobson2015}.
White noise parameter estimation is performed with the \textsc{ptmcmcsampler}~\citep{EllisvanHaasteren2019}.
Pulsar likelihoods are modelled using the code \textsc{enterprise}~\citep{EllisVallisneri2019} and linked to \textsc{pypolychord} with \textsc{bilby}~\citep{AshtonHubner2019}, using the code available at~\href{https://github.com/bvgoncharov/enterprise_warp}{github.com/bvgoncharov/enterprise\_warp}.



\section{Validating the general quasi-common noise model}
\label{sec:simqcp2}

We performed a study using a simulated data set to confirm the validity of the importance sampling method as well as the quasi-common spectrum process model, for which the common-spectrum process is a limiting case.
We note that the common-spectrum process hypothesis for a given pulsar is described by a likelihood from Equation~\ref{eq:signalstandard}, approximated with the importance sampling by Equation~\ref{eq:signal}.
This likelihood is then generalized in Equation~\ref{eq:noise} to represent the quasi-common spectrum process hypothesis.

\begin{figure*}
    \includegraphics[width=0.49\textwidth]{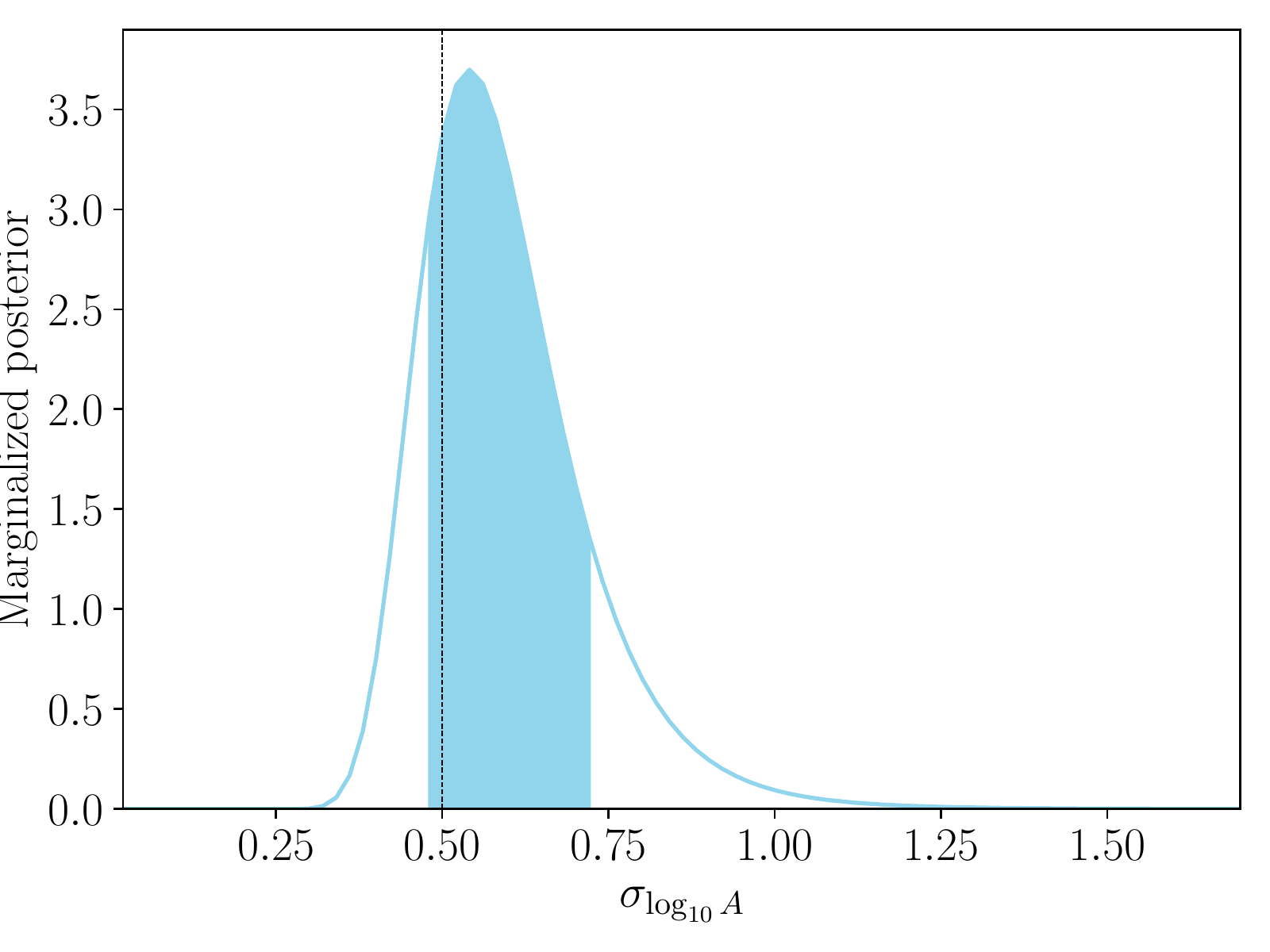}
    \includegraphics[width=0.49\textwidth]{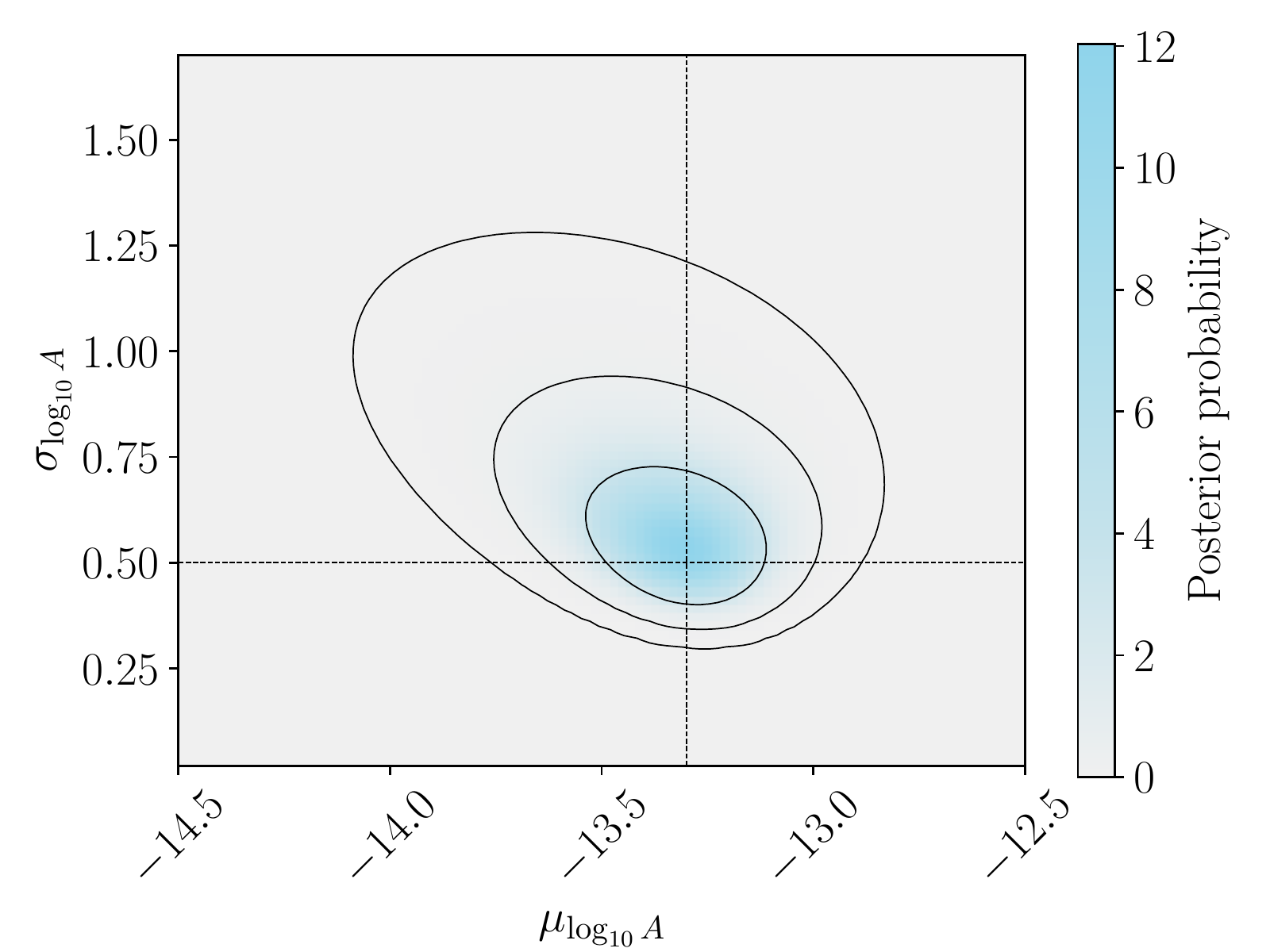}
    \caption{Searches for a quasi common-spectrum process  in a simulated data set.
    Left: marginalized posterior for $\sigma_{\log_{10}A}$. The value is inconsistent with zero, which means that the signal that would have been interpreted as the common-spectrum process with the standard methods is only a quasi-common spectrum process.
    Based on this result, the gravitational wave origin of the signal can be ruled out.
    The simulated value is represented by the vertical dotted line and cumulative $1$-$\sigma$ credible levels are represented by a shaded region.
    Right: full posterior on $\mu_{\log_{10}A}$ and $\sigma_{\log_{10}A}$. The vertical and the horizontal lines correspond to the simulated values. Three closed lines correspond to the standard $1$-, $2$- and $3$-$\sigma$ levels.}
    \label{fig:simsigma}
\end{figure*}

We simulated a data set with 26 pulsars, as in PPTA DR2, observed for 2555 days, with the pulse arrival time errors of $\sigma_\text{ToA}=0.1~\mu s$ and the white noise with variance proportional to $\sigma_\text{ToA}$.
We then simulate two red noise processes described by $\log_{10}A$ and $\gamma$, as we expect from the stochastic gravitational wave background and pulsar noise.
The simulations are based on 30 frequencies, which corresponds to the Fourier basis to model red noise in previous analyses.
One red noise process for each pulsar is drawn from the truncated Normal distribution $\mathcal{N}(\mu,\sigma)$ with $\mu_{\log_{10}A} = -16.3$, $\mu_{\gamma} = 5$, $\sigma_{\log_{10}A} = 2$, $\sigma_{\gamma} = 2$.
Another red noise process, the quasi-common noise, has a fixed $\gamma = 13/3$ and $\log_{10}A$ is drawn from the truncated Normal distribution with $\mu_{\log_{10}A} = -13.3$ and $\sigma_{\log_{10}A} = 0.5$.
Both Normal distributions are truncated to the edges of uniform priors used in~\cite{GoncharovShannon2021} to avoid edge effects.
The values were chosen so that pulsar-intrinsic red noise was below the detection threshold of some pulsars, as in PPTA DR2 where only 9 pulsars of 26 show evidence for spin noise.
Moreover, quasi-common noise realizations have simulated $A$ of the order of $10^{-14}$ - $10^{-13}$, as six realizations of the spin noise. The same number of pulsars in PPTA DR2 have the inferred common process amplitude and the spin noise amplitude of the same order of magnitude.
The results of parameter estimation on the simulated data set are presented in Figure~\ref{fig:simsigma}.
Both the $\mu_{\log_{10}A}$ and the $\sigma_{\log_{10}A}$ are consistent with the simulated values.
Note, we tested that this data set shows strong evidence for the common-spectrum process when using the priors from~\cite{GoncharovShannon2021}, and yet with our new generalized model we correctly infer that it is not the case because $\sigma_{\log_{10}A} \neq 0$.
More precisely, we find $\log \mathcal{B}^\text{CP}_\varnothing > 13.5$, where $\varnothing$ is the noise-only null hypothesis that includes only white and red noise, as per the simulations.

The noise in real data is more sophisticated than in the simulations, but it is not necessary to represent the whole complexity of the data to demonstrate the validity of the approach.
We defer exhaustive tests of the methodology to future work.
Our method is applicable to any data set under the assumption that physically relevant red noise processes (from interstellar propagation effects, instrumental noise, etc.) are separated from the common-spectrum process by including them in the models.
Moreover, we also find that the current simulation study is robust to the choice of priors for pulsar-intrinsic noise, with the only caveat being that the proposal likelihood should include all red noise terms in pulsars to allow parameters of both processes to be sampled.
We tested that the simulation works for a reduced case with only one red noise process per pulsar, correctly recovering position and width of a Gaussian distribution of pulsar red noise parameters.
We also trialed it for the case where the maximum probability density of the distributions of pulsar spin noise amplitudes ($\mu_{\log_{10}A} = -14.3$, $\sigma_{\log_{10}A} = 1.3$) and quasi-common noise amplitudes ($\mu_{\log_{10}A} = -13.8$, $\sigma_{\log_{10}A} = 0.4$) are similar and the distributions have a broader overlap.

\section{Derivation of the quasi-common process likelihood}
\label{sec:maths}

Whereas Equation~\ref{eq:signalstandard} models contributions from the common-spectrum process to be represented by one $A_\text{c}$ in all pulsars, we can generalize it to model different $A_\text{c}$ in pulsars. This way, assuming fixed $\gamma$, Equation~\ref{eq:signalstandard} transforms into
\begin{equation}\label{eq:step1}
    {\cal L}(\bm{d} | \bm{\theta}, A_{c,1}, ..., A_{c,N}) = \prod_{k=1}^N {\cal L}(\bm{d}_k | \bm{\theta}_k, A_{c,k}).
\end{equation}
Marginalizing over all possible values of $A_{c,k}$ in pulsars over the prior distribution these parameters are drawn from, which we model as a Gaussian distribution with $\mu$ and $\sigma$, the likelihood becomes
\begin{equation}\label{eq:step2}
    {\cal L}(\bm{d} | \bm{\theta}, \mu, \sigma) = \prod_{k=1}^N {\cal L}(\bm{d}_k | \bm{\theta}_k, \mu, \sigma) = \prod_{k=1}^N \int {\cal L}(\bm{d}_k | \bm{\theta}_k, A_{c,k}) \pi(A_{c,k}|\mu,\sigma) dA_{c,k}.
\end{equation}
Next, let us marginalize over ``nuisance'' parameters $\bm{\theta}$,
\begin{equation}\label{eq:step3}
    {\cal L}(\bm{d} | \mu, \sigma) = \prod_{k=1}^N \int {\cal L}(\bm{d}_k | \bm{\theta}_k, \mu, \sigma) \pi(\bm{\theta}_k) d\bm{\theta}_k.
\end{equation}
Let us then multiply the equation by unity and expand the unity via the proposal likelihood $\mathcal{L}(\bm{d}_k | \bm{\theta}_k)$,
\begin{equation}\label{eq:step4}
    {\cal L}(\bm{d} | \mu, \sigma) = \prod_{k=1}^N \int \frac{\mathcal{L}(\bm{d}_k | \bm{\theta}_k)}{\mathcal{L}(\bm{d}_k | \bm{\theta}_k)} {\cal L}(\bm{d}_k | \bm{\theta}_k, \mu, \sigma) \pi(\bm{\theta}_k) d\bm{\theta}_k.
\end{equation}
Representing one proposal likelihood through the posterior $\mathcal{P}(\bm{\theta}_k | \bm{d}_k)$, the evidence $\mathcal{Z}(\bm{d}_k)$, and the prior $\pi(\bm{\theta}_k)$, we obtain
\begin{equation}\label{eq:step5}
    {\cal L}(\bm{d} | \mu, \sigma) = \prod_{k=1}^N \int \mathcal{Z}(\bm{d}_k) \mathcal{P}(\bm{\theta}_k | \bm{d}_k) \bigg( \frac{{\cal L}(\bm{d}_k | \bm{\theta}_k, \mu, \sigma)}{\mathcal{L}(\bm{d}_k | \bm{\theta}_k)} \bigg) d\bm{\theta}_k,
\end{equation}
where the prior canceled out.
Next, we approximate the integral for a $k$'th pulsar with a sum over $n_k$ posterior samples, $\bm{\theta}_{k,j}$, providing
\begin{equation}\label{eq:step6}
    {\cal L}(\bm{d} | \mu, \sigma) = \prod_{k=1}^N \mathcal{Z}(\bm{d}_k) \frac{1}{n_k} \sum_{j=1}^{n_k} \frac{{\cal L}(\bm{d}_k | \bm{\theta}_{k,j}, \mu, \sigma)}{\mathcal{L}(\bm{d}_k | \bm{\theta}_{k,j})},
\end{equation}
where, omitting $\bm{\theta}_{k,j}$,
\begin{equation}\label{eq:step7}
    {\cal L}(\bm{d}_k | \mu, \sigma) = \int {\cal L}(\bm{d}_k | A_{c,k}) \pi(A_{c,k}|\mu,\sigma) dA_{c,k}.
\end{equation}
The approximation of an integral via the sum of posterior samples is explained for various applications in~\cite{HoggForeman-Mackey2018} and \cite{ThraneTalbot2019}.

\bibliography{mybib,separate}{}
\bibliographystyle{aasjournal}

\end{document}